%% file: main.tex
\documentclass[conference]{IEEEtran}
\IEEEoverridecommandlockouts

\usepackage[dvipsnames]{xcolor}
\usepackage{tikz}
\usetikzlibrary{shapes.geometric, arrows, positioning, shadows, decorations, decorations.text, shapes.misc, fit, arrows.meta, shadows}
\usepackage{amsmath,amssymb,amsfonts}
\usepackage{algorithmic}
\usepackage{graphicx}
\usepackage[footnotesize,labelfont=bf]{caption}
\usepackage{subcaption}
\usepackage{textcomp}
\usepackage[capitalize]{cleveref}

\usepackage[xindy, nonumberlist]{glossaries}
\makenoidxglossaries%
\loadglsentries{abbr}
\usepackage[backend=biber, style=ieee, citestyle=numeric-comp]{biblatex}
\AtBeginBibliography{\footnotesize}

\usepackage{xcolor}
\def\BibTeX{{\rm B\kern-.05em{\sc i\kern-.025em b}\kern-.08em
    T\kern-.1667em\lower.7ex\hbox{E}\kern-.125emX}}

\usepackage{xspace}
\newcommand{\lis}{\gls{cf}\xspace} %

\begin{document}

\title{Towards Practical Cell-Free 6G Network Deployments:\\ An Open-Source End-to-End Ray Tracing Simulator}

\author{
    \IEEEauthorblockN{
        William Tärneberg\textsuperscript{1},
        Aleksei Fedorov\textsuperscript{1},
        Gilles Callebaut\textsuperscript{2},
        Liesbet Van der Perre\textsuperscript{2},
        Emma Fitzgerald\textsuperscript{1}
    }
    \IEEEauthorblockA{
        \textsuperscript{1}Lund University, Department of Electrical and Information Technology, Sweden
    }
    \IEEEauthorblockA{
        \textsuperscript{2}KU Leuven, Department of Electrical Engineering, 9000 Ghent, Belgium
    }
    \thanks{6GTandem has received funding from the Smart Networks and Services Joint Undertaking (SNS JU) under the European Union’s Horizon Europe research and innovation programme under Grant Agreement No~101096302. 
    The REINDEER project has received funding from the European Union’s Horizon 2020 research and innovation programme under grant agreement No.~101013425. William is partially supported by the Wallenberg AI, Autonomous Systems and Software Program (WASP) funded by the Knut and Alice Wallenberg Foundation, the SEC4FACTORY project, funded by the Swedish Foundation for Strategic Research (SSF), and Imminence, a Celtic Next project funded by Sweden’s Innovation Agency (VINNOVA). The authors from Lund University are part of the Excellence Center at Linköping-Lund on Information Technology (ELLIIT), and the Nordic University Hub on Industrial IoT (HI2OT) funded by NordForsk. The work of Aleksei Fedorov within the SIVERT2 project is funded by Fordonsstrategisk forskning och innovation, FFI, and Sweden’s Innovation Agency. 
    }
}

\maketitle

\begin{abstract}
    The advent of \gls{6g} wireless communication marks a transformative era in technological connectivity, bringing forth challenges and opportunities alike. This paper unveils an innovative, open-source simulator, meticulously crafted for cell-free \gls{6g} wireless networks. This simulator is not just a tool but a gateway to the future, blending cutting-edge channel models with the simulation of both physical propagation effects and intricate system-level protocols. It stands at the forefront of technological advancement by integrating \gls{lis} and \gls{mimo} technologies, harnessing the power of the Unity game engine for efficient ray-tracing and \gls{gpu}-accelerated computations. The unparalleled flexibility in scenario configuration, coupled with its unique ability to dynamically simulate interactions across network layers, establishes this simulator as an indispensable asset in pioneering \gls{6g} systems' research and development.
\end{abstract}

\section{Introduction}
    The advent of \gls{6g} wireless networks heralds a transformative era in digital connectivity, characterized by ultra-low latency, high throughput, and enhanced reliability \cite{tataria20216g}. Such advancements demand innovative tools for simulation and analysis. In response, we present a novel, open-source simulator, specifically designed to tackle the complexities of cell-free \gls{6g} networks, particularly focusing on \lis systems as a key \gls{ran} architecture. Drawing upon research like Fedorov et al.~\cite{Fedorov2021}, our simulator (LuSim) aims to provide an in-depth analysis of network dynamics and channel behavior. The source code and documentation for LuSim are available on GitHub \cite{LuSim2023}.

    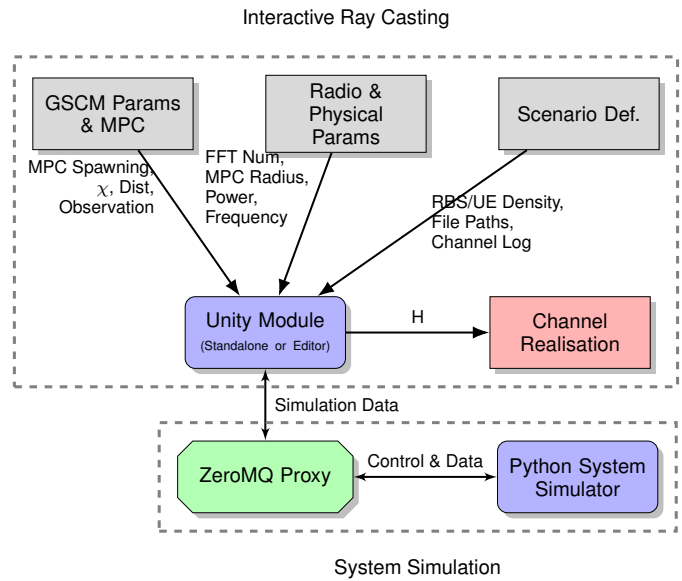
\begin{figure}[htbp]
        \centering
        \input{figures/architecture}
        
        \caption{System Architecture of the Wireless Network Simulator.}
        \label{fig:arch}
    \end{figure}

    Unlike traditional systems such as WiThRay~\cite{WiThRay}, QuaDRiGa~\cite{QuaDRiGa1, QuaDRiGa2}, and NVIDIA Sionna~\cite{SionnaRT}, our simulator integrates advanced \gls{gscm} and Exhaustive Ray Tracing methodologies. This integration allows for detailed and realistic simulations of wireless propagation in varied environments, from urban landscapes to indoor settings~\cite{VehicularMIMO, GSCMNonStationary, GSCMAdvantages, RayTracingStudy, RayTracingVisual}. By leveraging the Unity game engine, the simulator efficiently handles complex ray-casting tasks, making use of \gls{gpu} capabilities to serve the growing needs of researchers and engineers in this field.

    A distinctive feature of our tool is the incorporation of machine learning and artificial intelligence, which enables dynamic adaptation to diverse \gls{6g} network scenarios and user behaviors. This allows for enhanced performance analysis and optimization, offering insights beyond the reach of conventional simulation models.

    Our simulator's architecture is designed for flexibility and scalability, with externalized configurations and scenario definitions through JSON or YAML files. This modular approach supports a wide range of applications, including large-scale studies and sensitivity analyses. Furthermore, the simulator extends beyond its predecessors by including fully 3D indoor and cell-free scenarios, accommodating a large number of antennas and supporting studies on dynamic resource allocation.

    The cross-layer capabilities of our simulator facilitate a seamless interplay between the physical and application layers, enabling applications like resource orchestration and data synthesis for machine learning. This versatility ensures that the simulator remains a vital tool for exploring, developing, and deploying \lis-based \gls{6g} networks.

    In summary, our simulator represents a significant advancement in the simulation of \lis-based \gls{6g} networks. By combining the strengths of the Unity engine with a flexible and modular design, it offers a versatile platform for researchers and engineers, guiding the future of wireless communication technology.

    The development of simulators for wireless environments is crucial for understanding and improving communication systems. A notable contribution in this domain is WiThRay, a versatile ray-tracing (RT) simulator, designed for smart wireless environments~\cite{WiThRay}. WiThRay specializes in generating channel data for performance evaluations of complex communication techniques, including channel estimation/tracking, beamforming, and localization. It employs a novel RT algorithm that follows Fermat’s principle, reducing computational complexity, and includes features like scattering ray calibration and support for Reconfigurable Intelligent Surfaces (RIS) systems, making it highly relevant for 6G research.

    In comparison, other simulators like NVIDIA Sionna and QuaDRiGa~\cite{QuaDRiGa1, QuaDRiGa2} also make significant contributions to the field. NVIDIA Sionna, particularly noted for its comprehensive MIMO channel modeling capabilities, offers an in-depth analysis of wireless network performance. QuaDRiGa, with its focus on quasi-deterministic radio channel models, excels in large-scale parameter studies for mobile communication networks.

    Our simulator distinguishes itself by integrating a broad range of cutting-edge technologies and methodologies, enabling comprehensive analyses of complex 6G network scenarios. It leverages the strengths of \glspl{gscm}, widely adopted in academia and industry, for efficient modeling of complex urban and indoor environments. This approach is augmented with advanced features like dynamic environmental interaction, superior system-level simulation capabilities, and support for emerging technologies such as RIS and integrated sensing and communications.

    This unique combination of features in our simulator provides a versatile and adaptable platform, offering a deeper and more holistic understanding of cell-free 6G networks. Its ability to simulate intricate network dynamics and adapt to evolving wireless technologies positions it as a leading tool in the field, complementing and surpassing the capabilities of existing simulators like WiThRay, NVIDIA Sionna, and QuaDRiGa. The goal is to continuously evolve, integrating the latest advancements in wireless communication, to remain at the forefront of simulator development for 6G networks and beyond.

\section{Simulator Architecture}
    The architecture of our simulator is a testament to the evolving demands of cell-free \gls{6g} networks, uniquely designed to meet the intricate requirements of next-generation wireless communications. Central to this architecture is the Unity game engine, known for its exceptional graphical capabilities and efficient \gls{gpu} utilization. This choice empowers our simulator to provide accurate and dynamic modeling of the physical layer's challenges, particularly in the realms of ray casting and channel behavior simulation, surpassing traditional simulators like QuaDRiGa~\cite{QuaDRiGa1, QuaDRiGa2}, WiThRay~\cite{WiThRay}, and others in graphical fidelity and interactivity.
    
    The simulator's architecture is organized around three primary components, see \Cref{fig:arch}:
    
        \subsubsection{Configuration}
            Configuring the simulation is streamlined through JSON files, addressing aspects such as \textit{GSCM Parameters and \gls{mpc} Spawning}, \textit{Radio and Physical Parameters}, and \textit{Scenario Definition}. This modular approach allows for high flexibility and ease of customization, essential for diverse simulation scenarios, a feature that is uniquely advanced compared to the more rigid structures seen in simulators like WiThRay~\cite{WiThRay}.
    
        \subsubsection{Unity Module for \gls{phy} layer}
            Operating either as a standalone application or through the Unity editor, this module is at the forefront of simulating complex environments. It brings to life intricate urban and indoor settings with 3D models of \glspl{ue} and \glspl{bs}, which can be imported from tools like Blender~\cite{blender}. This module's capability to handle the actual simulation and physical layer, including the output of channel realizations, sets our simulator apart in terms of realism and detail, offering a more comprehensive solution than QuaDRiGa~\cite{QuaDRiGa1, QuaDRiGa2} and other existing systems.
    
        \subsubsection{System Simulator}
            Integrating seamlessly with the Unity module, this component takes over the control of time and specializes in simulating system orchestration scenarios like radio resource allocation. With its access to channel realizations and entity positions, and communication via UDP through a ZeroMQ proxy, the Python-based simulator enhances the overall functionality with features like a GUI and result plotting, offering a holistic tool for system-level analysis, a capability that is not fully explored in frameworks like NVIDIA Sionna~\cite{SionnaRT}.
    
    \subsection{Unity Components for GSCM}
        The Unity module in our simulator utilizes several components to effectively model GSCM in cell-free \gls{6g} environments:
        \begin{itemize}
            \item A detailed 3D scene representing the environment, like urban areas or indoor settings, with buildings and objects marked as reflecting surfaces.
            \item An active area defining the simulation scope and a traversable area for entity movement, leveraging Unity's mobility models or external control from the system simulator.
            \item Entities such as \glspl{ue} and \glspl{bs}, each equipped with one or more antennas, are placed in the scene. These entities can be added manually, populated randomly via the JSON scenario configuration, or controlled through the system simulator.
        \end{itemize}
        These components collectively enable the Unity module to simulate complex wireless interactions, providing a rich and interactive environment for studying \gls{6g} networks.

        This comprehensive architecture not only optimizes performance but also provides unmatched scalability and adaptability. It supports a wide array of scenarios, from complex urban landscapes to meticulously detailed indoor environments, making it an unparalleled tool in the study and development of advanced \gls{6g} systems.

\section{Unity Module - Interactive Ray Casting}
    Contrary to the conventional use of ray tracing in many simulators, our approach harnesses the advanced graphical capabilities of the Unity game engine for interactive ray casting, a key differentiator from other systems like WiThRay~\cite{WiThRay} and QuaDRiGa~\cite{QuaDRiGa1, QuaDRiGa2}. Ray casting, unlike ray tracing, involves projecting rays from a source to detect intersections with objects in the environment. This method is computationally less intensive and is particularly effective for modeling reflective interactions and basic propagation effects in complex environments, crucial for cell-free \gls{6g} networks~\cite{Liu2019, Yun2019, Kamaruddin2018, He2020}. Our method, drawing from the research of Aleksei Sivert et al.~\cite{Fedorov2021}, provides a unique approach to multi-path component modeling.

    This approach allows for real-time interactivity, enabling users to modify simulation parameters on the fly and immediately observe the impact on network performance and signal propagation. This level of dynamic interaction provides insights and analysis capabilities not commonly found in other simulators, which often rely on more computationally demanding ray tracing methods.

    \subsection{Initialisation of \gls{gscm} in Unity}
        The initialisation of \gls{gscm} in Unity involves several crucial steps, designed to prepare the system for both offline and real-time operation:
    \subsubsection{\gls{mpc} Distribution}
            The locations of scattering interactions, particularly along building facades, are key for accurate communication simulations. \glspl{mpc}, representing these scatterers, are uniformly distributed throughout the scene for each reflection order (1st, 2nd, and 3rd), with a unique density based on the COST model~\cite{gustafson2020cost}. This distribution enhances the realism of urban wireless communication scenarios and is automatically generated upon creating a scene~\cite{gustafson2020cost}.
    \subsubsection{\gls{mpc} Filtering}
            \glspl{mpc} that are inside objects or not visible to the entities (\glspl{ue} or \glspl{bs}) within the active area are removed using parallel ray casting, ensuring only relevant \glspl{mpc} are considered in the simulation.
    \subsubsection{Surface Association}
            Each \gls{mpc} is associated with the nearest surface, acquiring the normals of those surfaces. These normals, which may vary to simulate different scattering behaviors, play a critical role in determining how signals are reflected by the \gls{mpc}.
    \subsubsection{Path Gain Model}
            The path gain model for each \gls{mpc} accounts for distance dependence, interactions with scatterers, obstructions, diffraction around corners, angular dependence of scattering, and random large scale fading. The average path power gain for each \gls{mpc} is modeled using a classical log-distance power law, augmented with additional factors to represent these complex effects~\cite{gustafson2020cost}.
    \subsubsection{Parameter Estimation}
            A maximum-likelihood estimator is used to estimate the path power gain G0 and angular decay for each \gls{mpc}, aligning with the 3GPP Angular spread properties. These properties, which our simulator capably follows, are statistically derived from extensive experiments conducted by 3GPP active bodies. This alignment with 3GPP standards ensures that our modeling approach reflects real-world scenarios accurately. The estimation of G0 and $\xi$, while important, can be omitted if they are not directly utilized in the simulation. However, their inclusion offers depth to our model, allowing for a nuanced representation of the multi-path components' behavior in diverse environmental conditions~\cite{gustafson2020cost}.
    \subsubsection{Look-Up Table Generation}
            For higher order reflections (2nd and 3rd order), distance matrices are generated. These matrices help in determining the visibility of \glspl{mpc} to each other, which is essential in identifying possible signal paths using parallel ray casting.
    \subsubsection{Scene Representation}
\begin{figure}
            \centering
            \includegraphics[width=\columnwidth]{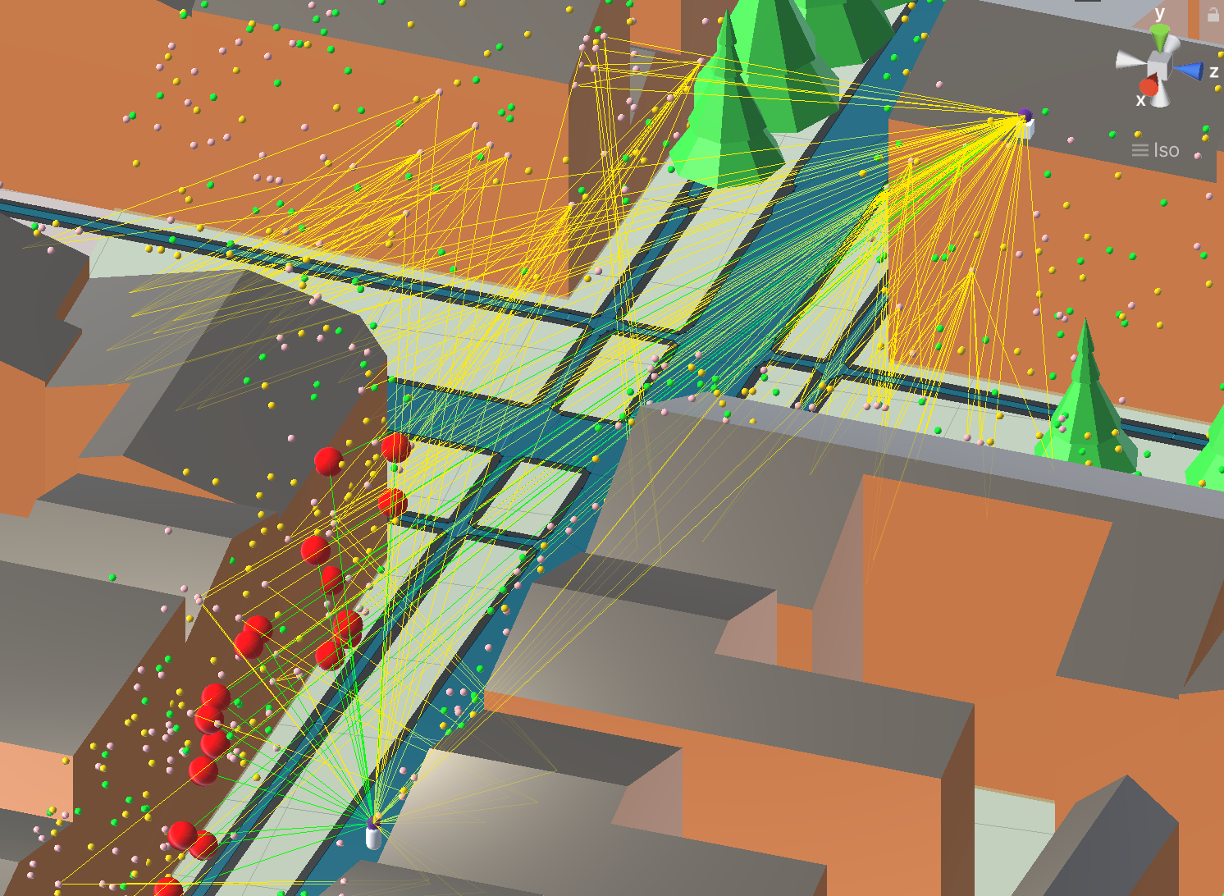}
            \caption{Visualization of \glspl{mpc} in an urban environment, illustrating both the spatial distribution and orientation of \glspl{mpc} (critical for signal propagation analysis) and the resultant signal paths. The scatter points (\glspl{mpc}) and their normals on surfaces demonstrate the complexity of the urban wireless environment. Colors indicate the order of paths: green for first order, yellow for second order, and red balls represent active MPCs. This comprehensive depiction provides insights into the intricacies of urban wireless propagation and the dynamics of signal paths in dense urban settings.}
            \label{fig:mpc_visualization_combined}
        \end{figure}

        The simulation environment is visually represented with color-coded \glspl{mpc} of different orders, providing a clear and intuitive understanding of the distribution and role of \glspl{mpc} in the simulated environment. Figure \ref{fig:mpc_visualization_combined} illustrates this visualization in an urban environment, highlighting the spatial distribution and orientation of MPCs along with the resultant signal paths. The color-coding, with green indicating first-order paths, yellow for second-order, and red balls representing active MPCs, enhances the scene's clarity and helps in comprehending complex urban wireless propagation dynamics.

        The static creation of \glspl{mpc} ensures spatial consistency across simulation runs, meaning that while the same \glspl{mpc} are present in each run, not all of them are visible depending on the specific scenario. 
        The MPC spawning process can also be saved and distributed for consistent simulation results, offering enhanced reproducibility and ease of simulation setup.

    \subsection{Run-time Channel Calculation}
        During run-time, the simulator calculates the channel as follows:
            \subsubsection{Instantaneous Path Power Gain} The simulator implements random shadow fading modeled as Gamma-distributed to represent fading around the distance-dependent mean path gain. This Gamma process, with its exponential autocorrelation, provides a realistic portrayal of time-variant power fading of individual multi-path components, enhancing the simulation's accuracy in depicting real-world channel conditions~\cite{gustafson2020cost}.
    
           \subsubsection{Model Validation} To ensure reliability, the model is validated against real-world measurement data. This validation process guarantees the accurate capture of channel gain, Doppler spread, and delay spread behaviors. It is especially crucial for intersections and urban scenarios not used in the parameter estimation phase, affirming the model's robustness and applicability in diverse environments~\cite{gustafson2020cost}.
    
            \subsubsection{Reflection Path Determination} Reflection paths are calculated using \glspl{mpc}, with the system accommodating various orders of reflections. For instance, a signal might first interact with a 2nd order \gls{mpc} and then with a 1st order \gls{mpc}, reflecting the complex nature of real-world signal propagation.
    
            \subsubsection{Path Length and Attenuation} The simulator allows setting a configurable maximum path length, measured in meters, to consider signal attenuation. This feature helps in limiting the number of potential paths and in simulating realistic signal strength reductions over distances.
    
            \subsubsection{Dynamic \gls{los} Determination} For moving \glspl{ue}, the simulator performs dynamic ray casting between each \gls{ue} and \gls{bs} to ascertain \gls{los} presence. This dynamic aspect is critical for simulating mobile environments where \gls{los} conditions can rapidly change.
    
            \subsubsection{Computational Efficiency} The simulation process is optimized for computational efficiency, employing techniques such as parallel processing and efficient path search algorithms. These optimizations are vital for real-time interactivity and for handling complex scenarios with multiple \glspl{ue} and \glspl{bs}.

    \subsection{Comparative Analysis of \glspl{gscm} and Exhaustive Ray Tracing in Wireless Simulators}
        In the domain of wireless network simulation for \gls{6g} networks, two primary methodologies are prominent for channel modeling: \glspl{gscm} and Exhaustive Ray Tracing. Our simulator strategically employs \glspl{gscm}, which provide a balance between computational efficiency and realistic environment simulation. This method, drawing insights from advanced models like the COST IRACON \glspl{gscm}, is apt for non-stationary environments and effectively captures the \gls{mimo} properties of channels~\cite{Liu2019, Yun2019}.
    
        The \glspl{gscm} in our simulator incorporate the distribution of scatterers in the environment, particularly along building facades, to model spatially dependent multi-path power. This approach accounts for building obstructions, diffraction around corners, and penetration losses in areas with trees, foliage, and other objects. The power fading for each MPC is modeled using a Gamma process, providing a stochastic yet accurate representation of the wireless channel~\cite{gustafson2020cost}.
    
        The path gain model for different MPCs includes effects such as distance dependence, losses due to interactions with scattering objects, and angular dependence of the scattering interaction. This comprehensive modeling approach ensures the accurate representation of real-world propagation environments, including non-stationary conditions prevalent in urban settings~\cite{gustafson2020cost}.
    
        Our implementation of \glspl{gscm} not only ensures spatial consistency of MPCs across simulation runs but also adheres to the \gls{dss} standard for channel storage~\cite{Call2311:Open}. While the same MPCs are statically created, their visibility varies in each run, depending on the specific scenario and environmental dynamics. This feature offers a balance between realism and computational efficiency, crucial for simulating complex 6G network environments. Additionally, the Unity module within our simulator can be utilized independently for detailed studies, providing flexibility in research and development applications~\cite{gustafson2020cost}.
    
        In contrast, Exhaustive Ray Tracing, as seen in simulators like those referenced in~\cite{Kamaruddin2018, He2020}, offers a deterministic approach for highly detailed environment-specific simulations but can be more resource-intensive. Our use of ray casting with Unity, combined with the versatility of \glspl{gscm}, enables us to offer detailed environmental modeling with greater computational efficiency. This combination allows our simulator to adapt to a wide range of scenarios, from complex urban layouts to specific indoor settings, providing a unique blend of adaptability, efficiency, and realism.
    
        In conclusion, the choice between \glspl{gscm} and exhaustive ray tracing or casting depends on the specific simulation needs. Our simulator’s integration of Unity's ray casting with \glspl{gscm} provides an ideal blend for advanced \gls{6g} network simulation, research, and development, outperforming other simulators in flexibility and operational efficiency.

\section{Python Module - System Simulator}\label{sec:python-sim}

    The second part of our simulator explores the complexities above the \gls{phy} layer, addressing the integration of intricate channel models with higher network layers. This integration is pivotal for examining joint and cross-layer algorithms, which are vital for optimizing the performance of \gls{6g} networks. Because the system-level simulator is decoupled from the channel model and interacts with it through a well-defined network protocol, researchers focusing on higher-layer problems can focus on implementing their solutions at the link layer or above, while still benefiting from the realistic channel modelling provided by the Unity module.

    \subsection{System Simulator Interaction}
        The system simulator interacts with the Unity module in two primary ways: i) through direct connection for dynamic simulation control, and ii) by reading out DSS-compliant channel measurements for detailed analysis. This Python-based simulator takes control of time from the Unity module, effectively orchestrating the Unity module's operations. Such an arrangement facilitates the orchestration of scenarios like dynamic resource allocation and network management. It accesses channel realizations and entity positions in the environment, communicating with the Unity module via \gls{udp} through a ZeroMQ proxy. Equipped with result plotting features, the system simulator serves as a comprehensive tool for system-level analysis and optimization, accommodating diverse simulation needs.

    \subsection{Discrete-Event Simulator}
        Our simulator leverages a high-level discrete-event Simulator~\cite{KeymeulenThesis} based on SimPy within the \gls{rw} infrastructure, offering a novel approach to network resource allocation. Each entity, such as a user or a \gls{csp}, manages its own behaviour and associated algorithms, while additional entities simulate the behaviour of system-wide logic such as centralized optimization algorithms.
    
    \subsection{Studying Novel Services}
         This infrastructure represents a novel distributed cell-free network, enhancing propagation and antenna array interactions. In this setup, the concept of federations—groupings of users and antennas—plays a crucial role~\cite{Call2207:Dynamic}. Managed through utility functions, these federations ensure fair allocation across the network, particularly in dynamic scenarios where user movements and changing requirements add layers of complexity. To do this, this module includes digital twins of the network infrastructure including energy, latency and other models. Using these, a subset of the resources, e.g., antennas, can be disabled to lower the energy expenditure of the network without compromising the \glspl{ue} requirements. 
    
         Next to wireless communication, the simulator is able to investigate \gls{wpt}. The channel gain from a dedicated channel or the Unity module is used to orchestrate the best federation to charge multiple devices in the most energy-efficient manner. Similarly, localisation and sensing is embedded in this module. It allows study of improving the precision and accuracy of the positioning, while simultaneously reducing the complexity by only selecting the best subset of radio resources available in the network.

    In summary, the above \gls{phy} layer of our simulator provides a detailed platform for simulating and analyzing various network management strategies. This part of the simulator is crucial for understanding and optimizing the deployment and operation of next-generation wireless networks, especially in the context of cell-free \gls{6g} network configurations.

    \glsresetall %
    
\section{Conclusions and Extensions}

    In conclusion, our simulator emerges as a significant milestone in the exploration and development of cell-free \gls{6g} wireless networks. It adeptly combines detailed physical layer modeling, utilizing advanced ray-tracing techniques powered by the Unity game engine, with an expansive system-level simulation framework. This synthesis of technologies enables a holistic examination of \gls{6g} networks, from the intricacies of signal propagation and interaction in diverse environments to complex network management strategies. 

    Our simulator's architecture, highlighted by its integration of \glspl{gscm}~\cite{VehicularMIMO, GSCMNonStationary, GSCMAdvantages} and Exhaustive Ray Tracing \cite{RayTracingStudy, RayTracingVisual}, offers a unique blend of realism and computational efficiency. This approach positions our simulator at the forefront of current research and development, surpassing other notable systems like WiThRay \cite{WiThRay}, QuaDRiGa \cite{QuaDRiGa1, QuaDRiGa2}, and even Sionna in terms of adaptability and detailed environmental modeling. Compared to Sionna, known for its comprehensive \gls{mimo} channel modeling, our simulator extends its capabilities by integrating more dynamic and varied environmental interactions and system-level complexities.

    Looking forward, we envision continual evolution and enhancement of the simulator. Our immediate focus is on integrating advanced features such as reconfigurable intelligent surfaces (RIS), which are gaining prominence in \gls{6g} research. This includes exploring integrated sensing and communications (ISAC) capabilities, which are pivotal for future intelligent networks. Additionally, we aim to refine dynamic resource allocation algorithms, ensuring efficient and adaptive network management in real-time scenarios.

    The goal is to solidify the simulator's status as an indispensable tool for researchers and engineers, facilitating breakthroughs in the field of wireless communications and contributing to the realization of the full potential of \gls{6g} networks. Our simulator will continually adapt, incorporating cutting-edge advancements and insights, thus ensuring its relevance and utility in the rapidly advancing domain of wireless technology. As the wireless communication landscape evolves, so too will our simulator, always staying a step ahead in the dynamic world of \gls{6g} wireless networks.

\printbibliography

\end{document}

%% file: abbr.tex
\newacronym{2d}{2D}{two-dimensional}
\newacronym{3d}{3D}{three-dimensional}
\newacronym{3gpp}{3GPP}{3rd generation partnership project}
\newacronym{4g}{4G}{fourth-generation}
\newacronym{5g}{5G}{fifth-generation}
\newacronym{6g}{6G}{sixth-generation}
\newacronym{abp}{ABP}{authentication by personalisation}
\newacronym{ac}{AC}{alternating current}
\newacronym{aclr}{ACLR}{adjacent channel leakage ratio}
\newacronym{adc}{ADC}{analog-to-digital converter}
\newacronym{adr}{ADR}{adaptive data rate}
\newacronym{afe}{AFE}{analog front end}
\newacronym{ag}{AG}{array gain}
\newacronym{ai}{AI}{artificial intelligence}
\newacronym{am}{AM}{amplitude modulation}
\newacronym{amam}{AM/AM}{amplitude modulation to amplitude modulation}
\newacronym{amp}{AMP}{approximate message passing}
\newacronym{ampm}{AM/PM}{amplitude modulation to phase modulation}
\newacronym{aoa}{AOA}{angle-of-arrival}
\newacronym{aod}{AOD}{angle-of-departure}
\newacronym{ap}{AP}{access point}
\newacronym{api}{API}{application program interface}
\newacronym{apu}{APU}{access point unit}
\newacronym{ar}{AR}{augmented reality}
\newacronym{arp}{ARP}{Antenna Reference Point}
\newacronym{asic}{ASIC}{application specific integrated circuit}
\newacronym{ask}{ASK}{amplitude-shift keying}
\newacronym{auv}{AUV}{autonomous underwater vehicle}
\newacronym{awgn}{AWGN}{additive white Gaussian noise}
\newacronym{baw}{BAW}{bulk acoustic wave}
\newacronym{bb}{BB}{base-band}
\newacronym{bc}{BC}{backscatter communication}
\newacronym{bd}{BD}{backscatter device}
\newacronym{ber}{BER}{bit error rate}
\newacronym{bf}{BF}{beamforming}
\newacronym{bldc}{BLDC}{brushless DC}
\newacronym{bom}{BOM}{bill of materials}
\newacronym{bs}{BS}{base station}
\newacronym{bw}{BW}{bandwidth}
\newacronym{ca}{CA}{carrier emitter}
\newacronym{cad}{CAD}{channel activity detection}
\newacronym{cars}{CARS}{calibration reference signal}
\newacronym{cbm}{CBM}{condition based maintenance}
\newacronym{cc}{CC}{constant current}
\newacronym{ccdf}{CCDF}{complementary cumulative distribution function}
\newacronym{ccnn}{CCNN}{circular convolutional neural network}
\newacronym{ccs}{CCS}{correlative channel sounder}
\newacronym{cdf}{CDF}{cumulative distribution function}
\newacronym{cdrx}{CDRX}{connected mode DRX}
\newacronym{ce}{CE}{coverage enhancement}
\newacronym{ced}{CED}{cumulative energy density}
\newacronym{cf}{CF}{cell-free}
\newacronym{cfmmimo}{CF-mMIMO}{cell-free massive MIMO}
\newacronym{cfo}{CFO}{carrier frequency offset}
\newacronym{cir}{CIR}{channel impulse response}
\newacronym{cmos}{CMOS}{complementary metal oxide semiconductor}
\newacronym{cost}{COST}{commercial off-the-shelf}
\newacronym{cots}{COTS}{commercial off-the-shelf}
\newacronym{cp}{CP}{cyclic prefix}
\newacronym{cpt}{CPT}{capacitive power transfer}
\newacronym{cpu}{CPU}{central-processing unit}
\newacronym{cqi}{CQI}{channel quality indicator}
\newacronym{cr}{CR}{coding rate}
\newacronym{crc}{CRC}{cyclic redundancy check}
\newacronym{crlb}{CRLB}{Cram\'er-Rao lower bound}
\newacronym{crs}{CRS}{cell reference signal}
\newacronym{cs}{CS}{compressed sensing}
\newacronym{cse}{CSE}{channel state estimation}
\newacronym{csi}{CSI}{channel state information}
\newacronym{csp}{CSP}{contact service point}
\newacronym{css}{CSS}{chirp spread spectrum}
\newacronym{cu}{CU}{central unit}
\newacronym{cv}{CV}{constant voltage}
\newacronym{cw}{CW}{continuous wave}
\newacronym{dac}{DAC}{digital-to-analog converter}
\newacronym{daq}{DAQ}{data acquisition system}
\newacronym{das}{DAS}{distributed antenna systems}
\newacronym{dc}{DC}{direct current}
\newacronym{dcc}{DCC}{dynamic cooperation clustering}
\newacronym{de}{DE}{drain efficiency}
\newacronym{dl}{DL}{downlink}
\newacronym{dlc}{DLC}{Distributed Laser Charging}
\newacronym{dli}{DLI}{direct link interference}
\newacronym{dm}{DM}{diffuse multipath}
\newacronym{dma}{DMA}{Direct Memory Access}
\newacronym{dmc}{DMC}{diffuse multipath component}
\newacronym{dmimo}{D-MIMO}{distributed MIMO}
\newacronym{doa}{DOA}{direction-of-arrival}
\newacronym{dpd}{DPD}{digital pre-distortion}
\newacronym{dpdk}{DPDK}{Data Plane Development Kit}
\newacronym{drx}{DRX}{Discontinuous Reception Mode}
\newacronym{dss}{DSS}{Dataset Storage Standard}
\newacronym{duc}{DUC}{digital up-converter}
\newacronym{e2e}{E2E}{end-to-end}
\newacronym{easa}{EASA}{European Union Aviation Safety Agency}
\newacronym{ec}{EC}{European Commission}
\newacronym{ecdf}{eCDF}{empirical cumulative distribution function}
\newacronym{ecsp}{ECSP}{edge computing service point}
\newacronym{edlc}{EDLC}{electrostatic double-layer capacitors}
\newacronym{edrx}{eDRX}{Extended Discontinuous Reception Mode}
\newacronym{ee}{EE}{energy efficiency}
\newacronym{egprs}{EGPRS}{Enhanced Data Rates for GSM Evolution}
\newacronym{eh}{EH}{energy harvesting}
\newacronym{eirp}{EIRP}{equivalent isotropically radiated power}
\newacronym{em}{EM}{electromagnetic}
\newacronym{embb}{eMBB}{enhanced Mobile Broadband}
\newacronym{en}{EN}{energy neutral}
\newacronym{end}{END}{energy neutral device}
\newacronym{eol}{EoL}{end of life}
\newacronym{epu}{EPU}{edge processing unit}
\newacronym{erp}{ERP}{equivalent radiated power}
\newacronym[plural=ESCs,firstplural=electronic speed controllers (ESCs)]{esc}{ESC}{electronic speed control}
\newacronym{esd}{ESD}{electrostatic discharge}
\newacronym{esl}{ESL}{electronic shelf label}
\newacronym{esr}{ESR}{equivalent series resistance}
\newacronym{etsi}{ETSI}{European Telecommunications Standards Institute}
\newacronym{evd}{EVD}{eigenvalue decomposition}
\newacronym{evm}{EVM}{Error Vector Magnitude}
\newacronym{fa}{FA}{federation anchor}
\newacronym{fair}{FAIR}{Findability, Accessibility, Interoperability, and Reuse of digital assets}
\newacronym{fd}{FD}{front-haul distance}
\newacronym{fdma}{FDMA}{frequency division multiple access}
\newacronym{fem}{FEM}{finite element analysis}
\newacronym{fembb}{feMBB}{further enhanced mobile broadband}
\newacronym{fft}{FFT}{fast Fourier transform}
\newacronym{fh}{FH}{fronthaul}
\newacronym{fim}{FIM}{Fisher information matrix}
\newacronym{fits}{FITS}{Flexible Image Transport System}
\newacronym{fom}{FoM}{figure of merit}
\newacronym{fpga}{FPGA}{field-programmable gate array}
\newacronym{fsk}{FSK}{frequency shift keying}
\newacronym{fss}{FSS}{frequency selective surface}
\newacronym{gb}{GB}{grant-based}
\newacronym{gf}{GF}{grant-free}
\newacronym{gnb}{gNB}{Next Generation Node B}
\newacronym{gnn}{GNN}{graph neural network}
\newacronym{gnss}{GNSS}{global navigation satellite system}
\newacronym{gpclk}{GPCLK}{general purpose clock}
\newacronym{gprs}{GPRS}{General Packet Radio Services}
\newacronym{gps}{GPS}{Global Positioning System}
\newacronym{gpu}{GPU}{graphical processing unit}
\newacronym{gscm}{GSCM}{geometry‐based stochastic model}
\newacronym{gsm}{GSM}{Global System for Mobile Communications}  %
\newacronym{gwp}{GWP}{Global Warming Potential}
\newacronym{harq}{HARQ}{hybrid automatic repeat request}
\newacronym{hat}{HAT}{hardware attached on top}
\newacronym{hcs}{HCS}{human-centric services}
\newacronym{hdf5}{HDF5}{Hierarchical Data Format version 5}
\newacronym{hpbm}{HPBM}{half power beam width}
\newacronym{HyMPRo}{HyMPRo}{Hybrid Multi-Path Routing algorithm}
\newacronym{i.i.d.}{i.i.d.}{independent and identically distributed}
\newacronym{ib}{IB}{in-band}
\newacronym{ibo}{IBO}{input back-off}
\newacronym{ic}{IC}{integrated circuit}
\newacronym{id}{ID}{information decoding}
\newacronym{if}{IF}{intermediate-frequency}
\newacronym{iid}{i.i.d.}{independently and identically distributed}
\newacronym{iis}{IIS}{integrated information system}
\newacronym{im}{IM}{intermodulation}
\newacronym{imu}{IMU}{inertial measurement unit}
\newacronym{io}{IO}{input/output}
\newacronym{iot}{IoT}{Internet of Things}
\newacronym{ipt}{IPT}{inductive power transfer}
\newacronym{ipy}{IPY}{Interventions per Year}
\newacronym{iq}{IQ}{in-phase and quadrature}
\newacronym{ir}{IR}{infrared}
\newacronym{ism}{ISM}{industrial, scientific and medical}
\newacronym{isp}{ISP}{internet service provider}
\newacronym{jfet}{JFET}{junction field effect transistor}
\newacronym{kpi}{KPI}{key performance indicator}
\newacronym{kvi}{KVI}{key value indicator}
\newacronym{larva}{LARVA}{LARge Virtual Array}
\newacronym{lca}{LCA}{life cycle assessment}
\newacronym{lco}{LCO}{lithium cobalt oxide}
\newacronym{ldo}{LDO}{Low-dropout voltage regulator}
\newacronym{ldpc}{LDPC}{low-density parity-check}
\newacronym{less}{LESS}{Low Energy Scheduler Solution}
\newacronym{lfp}{LFP}{lithium iron phosphate}
\newacronym{lic}{LIC}{lithium-ion capacitor}
\newacronym{lidar}{LiDAR}{light detection and ranging}
\newacronym{liion}{Li-ion}{lithium-ion}
\newacronym{lipo}{LiPo}{lithium polymer}
\newacronym{lis}{LIS}{large intelligent surface}
\newacronym{llh}{LLH}{log-likelihood}
\newacronym{lmmse}{LMMSE}{least minimum mean square error}
\newacronym{lmo}{LMO}{lithium ion manganese oxide}
\newacronym{lna}{LNA}{low-noise amplifier}
\newacronym{lo}{LO}{local oscillator}
\newacronym{lora}{LoRa}{long range}
\newacronym{lorawan}{LoRaWAN}{long-range wide-area network}
\newacronym{los}{LoS}{line-of-sight}
\newacronym{lp}{LP}{linear programming}
\newacronym{lpt}{LPT}{laser power transfer}
\newacronym{lpwa}{LPWA}{Low Power Wide Area}
\newacronym{lpwan}{LPWAN}{low-power wide-area network}
\newacronym{lqi}{LQI}{link quality indicator}
\newacronym{lrelu}{LReLU}{leaky rectified linear unit}
\newacronym{lrt}{LRT}{likelihood-ratio test}
\newacronym{ls}{LS}{least squares}
\newacronym{lsa}{LSA}{large synthetic array}
\newacronym{lsf}{LSF}{large-scale fading}
\newacronym{lsfc}{LSFC}{large-scale fading component}
\newacronym{ltc}{LTC}{lithium thionyl chloride}
\newacronym{lte}{LTE}{Long Term Evolution}
\newacronym{lti}{LTI}{linear time-invariant}
\newacronym{lto}{LTO}{lithium titanate}
\newacronym{m2m}{M2M}{machine to machine}
\newacronym{mac}{MAC}{Medium Access Control}
\newacronym{mate}{MATE}{millimeter-wave MIMO testbed}
\newacronym{mc}{MC}{Monte Carlo}
\newacronym{mcl}{MCL}{Maximum Coupling Loss}
\newacronym{mcs}{MCS}{modulation and coding scheme}
\newacronym{mcu}{MCU}{Microcontroller Unit}
\newacronym{mec}{MEC}{multi-access edge computing}
\newacronym{mems}{MEMS}{micro-electromechanical systems}
\newacronym{mf}{MF}{matched filter}
\newacronym{mimo}{MIMO}{multiple-input multiple-output}
\newacronym{miso}{MISO}{multiple-input single-output}
\newacronym{ml}{ML}{machine learning}
\newacronym{mlp}{MLP}{multilayer perceptron}
\newacronym{mmimo}{mMIMO}{massive MIMO}
\newacronym{mmse}{MMSE}{minimum mean square error}
\newacronym{mmtc}{mMTC}{massive machine-typed communication}
\newacronym{mmwave}{mmWave}{millimeter wave}
\newacronym{mn}{MN}{matching network}
\newacronym{mosfet}{MOSFET}{metal-oxide semiconductor field effect transistor}
\newacronym{mpc}{MPC}{multipath component}
\newacronym{mppt}{MPPT}{maximum power point tracking}
\newacronym{mr}{MR}{maximum ratio}
\newacronym{mrc}{MRC}{maximum ratio combining}
\newacronym{mrc_em}{MRC}{maximum ratio combining}
\newacronym{mrc_EM}{MRC}{Magnetic Resonance Coupling}
\newacronym{mrt}{MRT}{maximum ratio transmission}
\newacronym{mse}{MSE}{mean square error}
\newacronym{mtc}{MTC}{Machine-Type Communication}
\newacronym{multi-rat}{Multi-RAT}{multiple radio access technology}
\newacronym{music}{MUSIC}{MUltiple SIgnal Classification}
\newacronym{nb}{NB}{narrowband}
\newacronym{nbiot}{NB-IoT}{narrowband IoT}
\newacronym{nca}{NCA}{nickel cobalt aluminum}
\newacronym{netcdf}{NetCDF}{Network Common Data Form}
\newacronym{nfc}{NFC}{near-field communication}
\newacronym{ni}{NI}{National Instruments}
\newacronym{nicd}{NiCd}{nikkel cadmium}
\newacronym{nimh}{NiMH}{nikkel metal hydride}
\newacronym{nlos}{NLoS}{non-line-of-sight}
\newacronym{nmc}{NMC}{nickel manganese cobalt}
\newacronym{nmos}{nMOS}{n-channel metal-oxide semiconductor}
\newacronym{nn}{NN}{neural network}
\newacronym{nnls}{NNLS}{non-negative least squares}
\newacronym{noma}{NOMA}{non-orthogonal multiple access}
\newacronym{np}{NP}{Neyman-Pearson}
\newacronym{npbch}{NPBCH}{Narrowband Physical Broadcast Channel}
\newacronym{npss}{NPSS}{Narrow Band Primay Synchronization Signal}
\newacronym{nr}{NR}{New Radio}
\newacronym{nrs}{NRS}{Narrow Band Reference Signal}
\newacronym{nsss}{NSSS}{Narrowband Secondary Synchronization Signal}
\newacronym{ntp}{NTP}{network time protocol}
\newacronym{oai}{OAI}{OpenAirInterface} %
\newacronym{ofdm}{OFDM}{orthogonal frequency-division multiplexing}
\newacronym{ofdmim}{OFDM-IM}{OFDM with index modulation}
\newacronym{oob}{OOB}{out-of-band}
\newacronym{ook}{OOK}{on-off keying}
\newacronym{oran}{O-RAN}{open radio-access network}
\newacronym{os}{OS}{operating system}
\newacronym{ota}{OTA}{over-the-air}
\newacronym{otaa}{OTAA}{over-the-air authentication}
\newacronym{p1}{P1}{Phase 1}
\newacronym{p2}{P2}{Phase 2}
\newacronym{p2p}{P2P}{point-to-point}
\newacronym{pa}{PA}{power amplifier}
\newacronym{pae}{PAE}{power-added efficiency}
\newacronym{pana}{PanA}{Panel A}
\newacronym{panb}{PanB}{Panel B}
\newacronym{papr}{PAPR}{peak-to-average power ratio}
\newacronym{pc}{PC}{pilot count}
\newacronym{pcb}{PCB}{printed circuit board}
\newacronym{pcg}{PCG}{power consumption gain}
\newacronym{pcsi}{PCSI}{perfect channel state information}
\newacronym{pd}{PD}{powered device}
\newacronym{pdcch}{PDCCH}{physical downlink control channel}
\newacronym{pdf}{PDF}{probability density function}
\newacronym{pdp}{PDP}{power delay profile}
\newacronym{pdsch}{PDSCH}{physical downlink shared channel}
\newacronym{peb}{PEB}{positioning error bound}
\newacronym{per}{PER}{packet error rate}
\newacronym{pg}{PG}{path gain}
\newacronym{pgd}{PGD}{proximal gradient descent}
\newacronym{phy}{PHY}{physical}
\newacronym{pl}{PL}{path loss}
\newacronym{pll}{PLL}{phase-locked loop}
\newacronym{poe}{PoE}{power-over-Ethernet}
\newacronym{pps}{1PPS}{pulse per second}
\newacronym{prbs}{PRBs}{Physical Resource Blocks}
\newacronym{ps}{PS}{Processing System}
\newacronym{psd}{PSD}{power spectral density}
\newacronym{pse}{PSE}{power sourcing equipment}
\newacronym{psk}{PSK}{phase shift keying}
\newacronym{psm}{PSM}{power saving mode}
\newacronym{pss}{PSS}{primary synchronisation signal}
\newacronym{ptp}{PTP}{precision-time protocol}
\newacronym{ptrs}{PTRS}{Phase-Tracking Reference Signals}
\newacronym{ptw}{PTW}{paging time window}
\newacronym{pw}{PW}{planar wavefront}
\newacronym{pwm}{PWM}{pulse width modulation}
\newacronym{qam}{QAM}{quadrature amplitude modulation}
\newacronym{qos}{QoS}{quality-of-service}
\newacronym{quadriga}{QuaDRiGa}{QUAsi Deterministic RadIo channel GenerAtor}
\newacronym{ra}{RA}{Random Access}
\newacronym{ran}{RAN}{radio access network}
\newacronym{rar}{RAR}{Random Access Response}
\newacronym{rat}{RAT}{radio access technology}
\newacronym{rb}{Rb}{Rubidium}
\newacronym{rbs}{RBS}{radio base station}
\newacronym{rcs}{RCS}{radar cross section}
\newacronym{re}{RE}{radio element}
\newacronym{relu}{ReLU}{rectified linear unit}
\newacronym{rf}{RF}{radio frequency}
\newacronym{rfeh}{RFEH}{radio frequency energy harvesting}
\newacronym{rfic}{RFIC}{radio-frequency integrated circuit}
\newacronym{rfid}{RFID}{radio frequency identification}
\newacronym{rfpt}{RFPT}{radio frequency power transfer}
\newacronym{rfsoc}{RFSoC}{Radio Frequency System-on-Chip}
\newacronym{rir}{RIR}{room impulse response}
\newacronym{ris}{RIS}{reflective intelligent surface}%
\newacronym{rllmtc}{RLLMTC}{reliable low latency machine type communication}
\newacronym{rls}{RLS}{recursive least squares}
\newacronym{rms}{RMS}{root mean square}
\newacronym{rmse}{RMSE}{root-mean-square error}
\newacronym{rmt}{RMT}{random matrix theory}
\newacronym{rof}{RoF}{radio-over-fiber}
\newacronym{ros}{ROS}{robot operating system}
\newacronym{rpi}{RPi}{Raspberry Pi}
\newacronym{rrc}{RRC}{Radio Resource Connection}
\newacronym{rreq}{RREQ}{route request packet}
\newacronym{rsrp}{RSRP}{Reference Signals Received Power}
\newacronym{rss}{RSS}{received signal strength}
\newacronym{rssi}{RSSI}{received signal strength indicator}
\newacronym{rtc}{RTC}{real time clock}
\newacronym{rtk}{RTK}{real time kinematics}
\newacronym{rts}{RTS}{ray tracing simulator}
\newacronym{rv}{RV}{random variable}
\newacronym{rw}{RW}{RadioWeaves}
\newacronym{rx}{RX}{receiver}
\newacronym{rzf}{RZF}{regularized zero forcing}
\newacronym{s-parameter}{S-parameter}{scattering parameter}
\newacronym{sa}{SA}{synchronization anchor}
\newacronym{sar}{SAR}{specific absorption rate}
\newacronym{scfdma}{SCFDMA}{single-carrier frequency division multiple access}
\newacronym{sdg}{SDG}{Sustainable Development Goal}
\newacronym{sdm}{SDM}{sigma-delta modulator}
\newacronym{sdn}{SDN}{software-defined network}
\newacronym{sdof}{SDoF}{sigma-delta over fiber}
\newacronym{sdr}{SDR}{software-defined radio}
\newacronym{sf}{SF}{spreading factor}
\newacronym{sfn}{SFN}{single frequency network}
\newacronym{sfp}{SFP}{small form-factor pluggable}
\newacronym{SigMF}{SigMF}{Signal Metadata Format}
\newacronym{simo}{SIMO}{single-input multiple-output}
\newacronym{sinr}{SINR}{signal-to-interference-plus-noise ratio}
\newacronym{siso}{SISO}{single-input single-output}
\newacronym{slam}{SLAM}{simultaneous localization and mapping}
\newacronym{slc}{SLC}{spatial leakage suppression}
\newacronym{smc}{SMC}{specular multipath component}
\newacronym{smps}{SMPS}{switched mode power supply}
\newacronym{sndr}{SNDR}{signal-to-noise-and-distortion ratio}
\newacronym{snidr}{SNIDR}{signal-to-noise-and-interference-and-distortion ratio}
\newacronym{snr}{SNR}{signal-to-noise ratio}
\newacronym{soc}{SoC}{state of charge}
\newacronym{srd}{SRD}{short-range device}
\newacronym{ssb}{SSB}{synchronisation signal block}
\newacronym{ssd}{SSD}{solid state drive}
\newacronym{steam}{STEAM}{science, technology, engineering, the arts, and mathematics}
\newacronym{svd}{SVD}{singular value decomposition}
\newacronym{sw}{SW}{spherical wavefront}
\newacronym{tau}{TAU}{tracking area update}
\newacronym{tcer}{TCER}{transported to consumed energy ratio}
\newacronym{tcp}{TCP}{Transmission Control Protocol}
\newacronym{tdd}{TDD}{time division duplexing}
\newacronym{tdoa}{TDOA}{time-difference-of-arrival}
\newacronym{toa}{TOA}{time-of-arrival}
\newacronym{tof}{ToF}{time-of-flight}
\newacronym{tosm}{TOSM}{through-open-short-match}
\newacronym{trl}{TRL}{technology readyness level}
\newacronym{trp}{TRP}{Transmission Reception Point}
\newacronym{tsn}{TSN}{time-sensitive networking}
\newacronym{ttm}{TTM}{time to market}
\newacronym{ttn}{TTN}{The Things Network}
\newacronym{tx}{TX}{transmitter}
\newacronym{uav}{UAV}{unmanned aerial vehicle}
\newacronym{udp}{UDP}{User Datagram Protocol}
\newacronym{ue}{UE}{user equipment}
\newacronym{ugv}{UGV}{unmanned ground vehicle}
\newacronym{uhd}{UHD}{USRP hardware driver}
\newacronym{uhf}{UHF}{ultra-high frequency}
\newacronym{ul}{UL}{uplink}
\newacronym{ula}{ULA}{uniform linear array}
\newacronym{ummtc}{umMTC}{ultra massive machine type communication}
\newacronym{un}{UN}{United Nations}
\newacronym{upa}{UPA}{uniform planar array}
\newacronym{ura}{URA}{uniform rectangular array}
\newacronym{urllc}{URLLC}{ultra-reliable low-latency communications}
\newacronym{usrp}{USRP}{universal software radio peripheral}
\newacronym{uv}{UV}{unmanned vehicle}
\newacronym{uwb}{UWB}{ultrawideband}
\newacronym{v2v}{V2V}{vehicle-to-vehicle}
\newacronym{vep}{VEP}{virtual edge platform}
\newacronym{vlc}{VLC}{visible light communication}
\newacronym{vlp}{VLP}{visible light positioning}
\newacronym{vna}{VNA}{vector network analyzer}
\newacronym{votable}{VOTable}{Virtual Observatory Table}
\newacronym{vr}{VR}{virtual reality}
\newacronym{wb}{WB}{wideband}
\newacronym{wban}{WBAN}{wireless body area network}
\newacronym{wpt}{WPT}{wireless power transfer}
\newacronym{wr}{WR}{White Rabbit}
\newacronym{wrsn}{WRSN}{wireless rechargeable sensor network}
\newacronym{wsn}{WSN}{wireless sensor network}
\newacronym{xets}{XETS}{cross exponentially tapered slot}
\newacronym{xr}{XR}{extended reality}
\newacronym{z3ro}{Z3RO}{zero third-order distortion}
\newacronym{zf}{ZF}{zero-forcing}
\newacronym{zmcscg}{ZMCSCG}{zero mean circularly symmetric complex Gaussian}

\newacronym{srs}{SRS}{Sounding Reference Signal}

%% file: figures/architecture.tex
\resizebox{\columnwidth}{!}
{\begin{tikzpicture}[
    auto,
    file/.style={
        draw,
        rectangle,
        fill=gray!30,
        drop shadow,
        text width=2cm,
        align=center,
        minimum height=1cm,
        font=\sffamily\footnotesize
    },
    block/.style={
        draw,
        rectangle,
        rounded corners,
        fill=blue!30,
        drop shadow,
        text width=2cm,
        align=center,
        minimum height=1cm,
        font=\sffamily\footnotesize
    },
    proxy/.style={
        draw,
        chamfered rectangle,
        fill=green!30,
        drop shadow,
        text width=2cm,
        align=center,
        minimum height=1cm,
        font=\sffamily\footnotesize
    },
    line/.style={
        draw,
        -Latex,
        thick
    },
    line2/.style={
        draw,
        latex'-latex',
        thick
    },
    communication/.style={
        midway,
        font=\sffamily\scriptsize,
        fill=white, 
        fill opacity=0, 
        text opacity=1
    },
    section/.style={
        draw,
        rectangle,
        dashed,
        gray,
        very thick,
        inner sep=0.25cm
    }
]

\node (json1) [file] {GSCM Params \& MPC};
\node (json2) [file, right=of json1] {Radio \& Physical Params};
\node (json3) [file, right=of json2] {Scenario Def.};
\node (unity) [block, below=2cm of json2.south west, anchor=north] {Unity Module\\{\tiny (Standalone or Editor)}};
\node (proxy) [proxy, below=of unity] {ZeroMQ Proxy};
\node (python) [block, right=2cm of proxy] {Python System Simulator};
\node (output) [file, right=2cm of unity, fill=red!30] {Channel Realisation}; %

\draw[line] (json1) -- (unity) node[communication, near start, anchor=east, align=right] {MPC Spawning,\\$\chi$, Dist,\\Observation};
\draw[line] (json2) -- (unity) node[communication, near start, anchor=east, align=left] {FFT Num,\\MPC Radius,\\Power, \\Frequency};
\draw[line] (json3) -- (unity) node[communication, anchor=west, align=left] {RBS/UE Density,\\File Paths,\\Channel Log};
\draw[line2] (unity) -- (proxy) node[communication] {Simulation Data}; %
\draw[line2] (proxy) -- (python) node[communication] {Control \& Data}; %
\draw[line] (unity.east) -- (output.west) node[communication, anchor=south] {H};

\node[section] (raycasting) [fit=(json1) (json2) (json3) (unity) (output)] {};
\node[section] (systemsim) [fit=(proxy) (python)] {};

\node[above=0.25cm of raycasting.north, font=\sffamily\footnotesize] (raycastinglabel) {Interactive Ray Casting};
\node[below=0.25cm of systemsim.south, font=\sffamily\footnotesize] (systemsimlabel) {System Simulation};

\end{tikzpicture}}

%% file: references.bib
@inproceedings{Fedorov2021,
	title        = {{Implementation of spatially consistent channel models for real-time full stack C-ITS V2X simulations}},
	author       = {Aleksei Fedorov and Nikita Lyamin and Fredrik Tufvesson},
	year         = 2021,
	booktitle    = {Asilomar Conference on Signals, Systems, and Computers},
	organization = {IEEE}
}

@article{GSCMAdvantages,
	title        = {{A Geometry-Based Stochastic Channel Model for MIMO Mobile-to-Mobile Communications}},
	author       = {Wang, Cheng-Xiang and Cheng, Xiang and Laurenson, David I and Salous, Sana and Vasilakos, Athanasios V},
	year         = 2008,
	journal      = {Transactions on Vehicular Technology},
	publisher    = {IEEE}
}

@article{GSCMNonStationary,
	title        = {{Evaluation of the Spatial Consistency Feature in the 3GPP Geometry-Based Stochastic Channel Model}},
	author       = {Kolmonen, Veli-Matti and Vehkapera, Mikko and Jamsa, Tuomas and Molisch, Andreas F},
	year         = 2011,
	journal      = {Transactions on Antennas and Propagation},
	publisher    = {IEEE}
}

@article{gustafson2020cost,
	title        = {{The COST IRACON Geometry-Based Stochastic Channel Model for Vehicle-to-Vehicle Communication in Intersections}},
	author       = {Gustafson, Carl and Mahler, Kim and Bolin, David and Tufvesson, Fredrik},
	year         = 2020,
	journal      = {Transactions on Vehicular Technology},
	publisher    = {IEEE}
}

@article{He2020,
	title        = {{The Design and Applications of High-Performance Ray-Tracing Simulation Platform for 5G and Beyond Wireless Communications: A Tutorial}},
	author       = {Danping He and others},
	year         = 2020,
	journal      = {Transactions},
	publisher    = {IEEE}
}

@article{Kamaruddin2018,
	title        = {{A Comprehensive Review of Efficient Ray-Tracing Techniques for Wireless Communication}},
	author       = {Mohd Nazeri Kamaruddin and others},
	year         = 2018,
	journal      = {International Journal on Communications Antenna and Propagation (IRECAP)}
}

@mastersthesis{KeymeulenThesis,
	title        = {{High-Level Simulator of Federation Orchestration}},
	author       = {Toon Keymeulen},
	year         = 2022,
	school       = {Lund University}
}

@article{Liu2019,
	title        = {{Optimization of Multi-UAV-Aided Wireless Networking Over a Ray-Tracing Channel Model}},
	author       = {An Liu and Vincent K. N. Lau},
	year         = 2019,
	journal      = {Transactions},
	publisher    = {IEEE}
}

@article{QuaDRiGa1,
	title        = {{QuaDRiGa: A 3-D Multi-Cell Channel Model With Time Evolution for Enabling Virtual Field Trials}},
	author       = {Other Authors},
	year         = {Year},
	journal      = {Transactions on Antennas and Propagation},
	publisher    = {IEEE}
}

@article{QuaDRiGa2,
	title        = {{Investigation and Comparison of QuaDRiGa, NYUSIM and MG5G Channel Models for 5G Wireless Communications}},
	author       = {He, Yaping and Zhang, Yang and Zhang, Jin and Pang, Lihua and Chen, Yijian and Ren, Guangliang},
	year         = 2020,
	booktitle    = {Vehicular Technology Conference (VTC2020-Fall)},
	publisher    = {IEEE}
}

@article{RayTracingStudy,
	title        = {{Ray Tracing for Radio Propagation Modeling: Principles and Applications}},
	author       = {Degli-Esposti, Vittorio and Fuschini, Franco and Vitucci, Enrico M and Falciasecca, Gabriele},
	year         = 2015,
	journal      = {Access},
	publisher    = {IEEE}
}

@article{RayTracingVisual,
	title        = {{Exhaustive Ray Tracing for 6G Topics such as RIS or Integrated Sensing and Communications}},
	author       = {Smith, John and Johnson, Emily},
	year         = 2022,
	journal      = {Journal of Advanced Communications},
	publisher    = {Advanced Communications Press}
}

@article{SionnaRT,
	title        = {{Sionna RT: Differentiable Ray Tracing for Radio Propagation Modeling}},
	author       = {Hoydis, Jakob and Aoudia, Fayc\c{}al Ait and Cammerer, Sebastian and others},
	year         = 2023,
	journal      = {arXiv preprint arXiv:2303.11103}
}

@article{VehicularMIMO,
	title        = {{A Geometry-Based Stochastic MIMO Model for Vehicle-to-Vehicle Communications}},
	author       = {Karedal, Johan and Czink, Nicolai and Paier, Alexander and Tufvesson, Fredrik and Molisch, Andreas F},
	year         = 2011,
	journal      = {Transactions on Vehicular Technology},
	publisher    = {IEEE}
}

@article{WiThRay,
	title        = {{WiThRay: A Versatile Ray-Tracing Simulator for Smart Wireless Environments}},
	author       = {Choi, Hyuckjin and Oh, Jaeky and Chung, Jaehoon and Alexandropoulos, George C. and Choi, Junil},
	year         = 2023,
	journal      = {Access},
	publisher    = {IEEE}
}

@article{Yun2019,
	title        = {{An Integrated Method of Ray Tracing and Genetic Algorithm for Optimizing Coverage in Indoor Wireless Networks}},
	author       = {Zhengqing Yun and Sungkyun Lim and Magdy F. Iskander},
	year         = 2019,
	journal      = {Transactions},
	publisher    = {IEEE}
}

@misc{blender,
	title        = {Blender},
	author       = {{Blender Foundation}},
	year         = 2023,
	url          = {https://www.blender.org},
	note         = {[Software]},
	version      = {4.0}
}

@article{tataria20216g,
	title        = {6G wireless systems: Vision, requirements, challenges, insights, and opportunities},
	author       = {Tataria, Harsh and Shafi, Mansoor and Molisch, Andreas F and Dohler, Mischa and Sj{\"o}land, Henrik and Tufvesson, Fredrik},
	year         = 2021,
	journal      = {Proceedings of the IEEE},
	publisher    = {IEEE}
}

@inproceedings{Call2207:Dynamic,
	title        = {Dynamic Federations for {6G} {Cell-Free} Networking: Concepts and Terminology},
	author       = {Gilles Callebaut and William {T{\"a}rneberg} and Liesbet {Van der Perre} and Emma Fitzgerald},
	year         = 2022,
	booktitle    = {International Workshop on Signal Processing Advances in Wireless Communication (SPAWC)},
	publisher    = {IEEE}
}

@inproceedings{Call2311:Open,
	title        = {An Open Dataset Storage Standard for {6G} Testbeds},
	author       = {Gilles Callebaut and Michiel Sandra and Christian Nelson and Thomas Wilding and Daan Delabie and Benjamin J. B. Deutschmann and William {T{\"a}rneberg} and Emma Fitzgerald and Anders Johansson and Liesbet {Van der Perre}},
	year         = 2023,
	booktitle    = {Conference on Antenna Measurements and Applications (CAMA)},
	publisher    = {IEEE}
}

@misc{LuSim2023,
  author = {6G-Testbeds},
  title = {LuSim: An Open-Source Simulator for Cell-Free 6G Networks},
  year = {2023},
  howpublished = {\url{https://github.com/6G-Testbeds/LuSim}}
}
